\documentclass{ws-ijmpc}

\begin{document}

\catchline{}{}{}{}{}

\title{PRISONER'S DILEMMA IN ONE-DIMENSIONAL CELLULAR AUTOMATA: VISUALIZATION OF EVOLUTIONARY PATTERNS}

\author{MARCELO ALVES PEREIRA$^\dagger$, ALEXANDRE SOUTO MARTINEZ$^\ddagger$}
\address{
Departamento de F\'{\i}sica e Matem\'{a}tica \\
Faculdade de Filosofia Ci\^encias e Letras de Ribeir\~{a}o Preto \\
Universidade de S\~ao Paulo \\
Av. Bandeirantes, 3900, 14040-901 \\
Ribeir\~{a}o Preto, SP, Brazil \\
$^\dagger$marceloapereira@usp.br \\
$^\ddagger$asmartinez@usp.br}

\author{AQUINO LAURI ESP\'{I}NDOLA}
\address{
Departamento de F\'{\i}sica e Matem\'{a}tica \\
Faculdade de Filosofia Ci\^encias e Letras de Ribeir\~{a}o Preto \\
Departamento de Medicina Social \\
Faculdade de Medicina de Ribeir\~ao Preto \\
Universidade de S\~ao Paulo\\
Av. Bandeirantes, 3900, 14049-900 \\
Ribeir\~{a}o Preto, SP, Brazil \\
aquinoespindola@usp.br}

\maketitle

\begin{history}
\received{Day August 2007}
\revised{Day Month 2007}
\end{history}

\begin{abstract}
The spatial Prisoner's Dilemma is a prototype model to show the emergence of cooperation in very competitive environments.
It considers players, at site of lattices, that can either cooperate or defect when playing the Prisoner's Dilemma with other $z$ players.
This model presents a rich phase diagram.
Here we consider players in cells of one-dimensional cellular automata.
Each player interacts with other $z$ players.
This geometry allows us to vary, in a simple manner, the number of neighbors ranging from one up to the lattice size, including self-interaction.
This approach has multiple advantages.
It is simple to implement numerically and we are able to retrieve all the previous results found in the previously considered lattices, with a faster convergence to stationary values.
More remarkable, it permits us to keep track of the spatio-temporal evolution of each player of the automaton.
Giving rise to interesting patterns.
These patterns allow the interpretation of cooperation/defection clusters as particles, which can be absorbed and collided among themselves.
The presented approach represents a new paradigm to study the emergence and maintenance of cooperation in the spatial Prisoner's Dilemma.

\end{abstract}

\ccode{PACS Nos.: 02.50.Le, 07.05.Tp}

\section{Introduction} \label{introduction}
Games amuse mankind since emerging of the first civilizations.
Besides amusement, they are a branch of Mathematics known as Game Theory, consolidated by John von Neumann.\cite{neumann1947}
The main purpose of such theory is the determination of strategies to get the maximum returns in situations where multiple rational players have the same aim.
The most prominent game, due to the cooperation emergence among competitive rational players, is the \emph{Prisoner's Dilemma} (PD), originally framed by Merrill Flood and Melvin Dresher in 1950.\cite{dresher}
The game formalization with prison sentence payoffs and the name ``Prisoner's Dilemma'' is due to Albert W. Tucker, when he wanted to make the ideas of Flood and Dresher more accessible to an audience at Stanford University in the same year.\cite{poundstone1992}

There are several problems that can be modeled using the PD game.
In real life, there are many situations of conflict, i.e., one person trying to reach his/her personal aim is incompatible to collective aims.
Because PD is a conflict situation between two players, it is possible to make an analogy between the game and real life, e.g., in Politics (Sociophysics)\cite{stauffer2004}, Economics (Econophysics)\cite{bouchaud2002,anteneodo2002}, and Biology.\cite{turner1999}

In the classical version of PD, two players can either cooperate or defect.
Under mutual cooperation players get a payoff $R$ (reward), otherwise if they are defectors, the payoff is $P$ (punishment).
When a player cooperates and the other defects, they get $S$ (sucker) and $T$ (temptation), respectively.
These payoff values must satisfy the inequalities:
$T > R > P > S$ and $T + S < 2R$.
The former relationship assures the existence of the dilemma, whereas the latter prevents that a couple of players, that alternate between cooperation and defection, get the same payoff or the payoff be higher than a cooperative couple payoff.

In a single round game, the best choice for a rational player is to defect.
Defection assures the largest payoff independently of the other player's decision (Nash equilibrium).
In successive rounds, one has the Iterated Prisoner Dilemma (IPD).
In this case, the best choice is not necessarily the defection, it is convenient only to retaliate a previous non-cooperative play.
The IPD game became popular with the computer tournament proposed by Axelrod\cite{axelrod1981,axelrod1984}.
This tournament intended to compare the strategies.
It turned out, that a simple strategy, with only one time step memory, called tit-for-tat (TFT), was by far the most stable one.

Nowak and May\cite{nowak1992} have shown the emergence of cooperation between players with strategies with no memory, in the presence of spatial structure.
This version of the game is known as Spatial Prisoner's Dilemma (SPD).
This is a simple, purely deterministic and spatial version of the classical PD, but it may generate chaotically changing spatial-temporal patterns, where cooperators and defectors coexist, with cooperators proportion oscillating indefinitely.
This effect is present when each player interacts with the nearest neighbors.
Adding interaction with the next nearest neighbors (corresponding to the chess king's move) the spatial patterns are smoother.
The outcome is the final proportion of cooperators and defectors, and in the chaotic phase, it depends on the initial configuration and on the magnitude of the parameter $T$.
The spatial structure is the factor that explains the maintenance of cooperation.\cite{nowak1992}
Here, cooperative/defective clusters are formed and evolve dynamically.
The invasions of cooperative clusters by defectors, and vice-versa, occur on the borders of the clusters, for this reason it is important to understand border effects.
Furthermore, connectivity among players also plays an important role in the dynamics of the clusters.\cite{duran2005}

The Prisoner's Dilemma has been studied in lattices with different topologies such as square lattice\cite{nowak1992}, graphs\cite{nowak2005} and also in complex networks as random graphs\cite{duran2005}, scale-free networks\cite{wu2007}, small-world networks.\cite{abramson2001}
However, in the simplest lattice topology, i.e., one-dimensional lattice, studies have not been carried out, possibly due its apparent simplicity and also because of simple behavior when only next nearest neighbors are considered.
Beyond topologies, the mobility of players can also be considered.\cite{arenzon2007}

We study the PD using the one-dimensional lattice with a variable number of interacting neighbors.
This geometry allows us to vary, in a simple manner, the number of neighbors ranging from one, up to the lattice size.
In this lattice, the cooperative/defective clusters are localized.
It is simpler to understand the invasions, which can occur only on the two borders of the one-dimensional cooperative/defective clusters.
The computational implementation of PD in the one-dimensional case is simpler than in a square lattice or honeycomb lattice, for instance, and it requires less computational time to process the numerical code.

Another relevant result of the PD implementation in one-dimensional automata is the emergence of patterns similar to those observed in the cellular automata theory.\cite{wolfram1983}
There is also the possibility of understanding the role of local interaction easily, furnishing a new perspective in the study of the PD.
There are four basic qualitative classes of behavior that empirically characterize the cellular automata, whose time evolution generates different outcomes.
Class 1, leads to a homogeneous state, all the cells get the same state.
Class 2, leads to a set of stable or periodic structures that are separated and simple.
Class 3, leads to the formation of chaotic pattern.
Class 4, leads to complex structures, sometimes long-lived.\cite{wolfram1983}

In this paper, we present the IPD in cellular automata where the number of interaction among the players can vary.
After introducing the model in Section \ref{themodel}, we show the equivalence with previous models.
Further, we explore the fact that we are able to keep track of time evolution to study the invasion dynamics of cooperative clusters by defectors.
The patterns found are presented in Section \ref{resultados}.
Final remarks are presented in Section \ref{conclusao}.

\section{The Model} \label{themodel}
In our model, we consider cellular automata in a one-dimensional lattice, with $L$ cells.
Each cell (grid position) corresponds to one player with two possible states: $\theta = 1$ for cooperator and $\theta = 0$ for defector.
The automata have no empty cells, so cooperators and defectors have the proportion $\rho_c(t) = (1/L) \sum_{i = 1}^{L} \theta_i(t)$ and $\rho_d(t)$, respectively, which implies $\rho_c(t) + \rho_d(t) = 1$.

The initial proportion of cooperators, $\rho_c(0) \equiv \rho_0$, with $0 < \rho_0 < 1$, is an important parameter in the problem.
The distribution of these $\rho_0$ cooperators in the lattice is random with an uniform distribution.
This initial distribution is the only stochastic variable in the model.

The number of players that interact with player $i$ is given by $z = (1, 2, \ldots, L)$.
The neighborhood $z$ can be symmetric or asymmetric.
For a symmetric neighborhood, if $z$ is even, there are $\alpha = z/2$ adjacent interacting players to the right hand side and $\alpha$ to the left hand side of the player $i$.
If $z$ is odd, each side has $\alpha = (z-1)/2$ players and player $i$ interacts against his/her own state (self-interaction).
For an asymmetric neighborhood, if $z$ is even, there are $\alpha_r$ adjacent interacting players to the right hand side and $\alpha_l$ to the left hand side of the player $i$, where $\alpha_r + \alpha_l = z$.
If $z$ is odd, moreover the $\alpha_r$ and $\alpha_l$ neighbors, the player $i$ interacts against his/her own state (self-interaction), where $\alpha_r + \alpha_l = z - 1$.
Nowak and May\cite{nowak1993} argue that the self-interaction makes sense if several animals (a family) or molecules may occupy a single patch.
Most of the studies do not show clearly how self-interaction works.
Despite the fact of the short time and asymptotic behavior of $\rho_c$ are similar with or without self-interaction, the intermediate temporal behavior is different as shown in Refs.

The player's conflict between cooperating or defecting, is well defined by the payoff parameters.
Nowak and May\cite{nowak1992} used modified Tucker's values, i.e., $R=1$, $S=P=0$, leaving only one free parameter, the temptation $T$.
The conditions $T > R > P > S$ and $T + S < 2R$ have been relaxed ($P = S$, and when $T = 2$, $T + S = 2R$) without any harm.
Due to this new setting, the conflict range is set to $1 < T < 2$.
The macroscopic regimes also depend on the $z$ values.

Consider the players $i$ and $j$ playing PD in the cellular automaton.
Player $i$ has a payoff due the interaction with player $j$ given by

\begin{equation} \label{eq_payoff}
    g_{\theta_i,\theta_j} = \theta_i \theta_j + T(1-\theta_i \theta_j)\theta_j.
\end{equation}
The total payoff, $P_i$, of the player $i$ is:

\begin{equation} \label{eq_totalpayoff}
    P_i = \sum_{j=1}^{z} g_{\theta_i,\theta_j}.
\end{equation}

We stress that the model dynamics is totally deterministic.
Starting from the left hand side to the opposite one, player $i$ will compare $P_i$ to $P_k$, where $P_k$ is the payoff of the $k$-th neighbor, where $k = (1, 2, \ldots, z)$.
As mentioned before, only if $z$ is odd, there is an extra payoff component $g_{\theta_i,\theta_i}$ due the self-interaction.
During the payoff comparison process, two situations can occur: i) if $P_i \geq \max(P_k)$, player $i$ will keep the current state; or ii) $P_i < \max(P_k)$, then player $i$ will adopt the state of the player $j$, which is the one with highest payoff in the $k$ set.
The states of the players are updated synchronously until the system reaches a stationary or a dynamical equilibrium state.
The process described here is known as Darwinian Evolutionary Strategy.
There are also other Evolutionary Strategies that can be adopted, such as Pavlovian.\cite{fort2005}

Notice the dependence of $\rho_c(t,T,\rho_0,z)$ on time, temptation, initial proportion of cooperators, and number of interacting players.
The dependence of $\rho_c$ as function of $\rho_0$ is frequently neglected.
The dependence on $z$ is also neglected due to fixed lattice restriction.
When system reaches a steady state, one can observe the asymptotic proportion of cooperators, $\rho_\infty(T,\rho_0,z)$, which represents the final phase of the system for the set of parameters $(T,\rho_0,z)$.
According to Ref. \refcite{nowak1992}, three different qualitative regimes are well defined, as function of $T$, for this system with $z = 9$.
For lower [upper] values of $T$, $1 < T < 4/3$ [$3/2 < T < 2$], the asymptotic values for the proportion of cooperators, $\rho_\infty$ is stationary or slightly periodical and became a majority ($\rho_\infty > 0.5$) [minority ($\rho_\infty < 0.5$)].
For intermediate values of $T$, ($4/3 < T < 3/2)$, $\rho_\infty(T)$ is non-stationary, even rendering spatial-temporal chaos because it is strongly dependent on the initial configurations.\cite{nowak1992}

The strong dependence of $\rho_\infty$ on $z$ can be understood observing the microscopic states, and counting the number interacting cooperators in the neighborhood $c$ of a given player, where $0 \leq c \leq z$.
From Eq. (\ref{eq_payoff}) one can see that when player $i$ interacts with $c_i$ cooperators of the neighborhood of $z$ players, his/her payoff is\cite{duran2005}:

\begin{equation} \label{eq_totalpayoffviz}
    P_{i}^{(c_i)}(\theta_i) = [T-(T-1)\theta_i]c_i.
\end{equation}

Some useful relations follow immediately from equation Eq. (\ref{eq_totalpayoffviz}): for a cooperator $P_{i}^{(c_i)}(1) = c$, whereas for a defector $P_{i}^{(c_i)}(0) = cT$.
For $T > 1$, $P_{i}^{(c_i)}(0) > P_{i}^{(c_i)}(1)$ and $P_{i}^{(c)}(\theta) \geq P_{i}^{(c - 1)}(\theta)$.
Transitions in $\rho_\infty(T)$ occur when temptation on the border of the cooperative cluster cross a threshold value.
In the conflict range, these transitions, $1 < T < 2$, are controlled by\cite{duran2005}:

\begin{equation} \label{eq_Tc}
    T_c(n,m) = \frac{z-n}{z-n-m},
\end{equation}
where $0 \leq n < z$ and $1 \leq m \leq \mbox{int}[(z-n-1)/2)]$ are integers.

\section{Results}\label{resultados}
In our simulations, we have used one-dimensional cellular automata with $L = 1,000$ cells in which $\rho_0$ are cooperators and the remaining cells are set as defectors.
To avoid dependence on the initial configurations, the values of the asymptotic proportion of cooperators, $\rho_\infty$, are mean values of ensembles of 1,000 realizations.
The comparison between preliminary results obtained with $L = 1,000$ and $L = 10,000$ has shown that $L = 1,000$ is good enough to study the behavior of the $\rho_c$.
From now on, all the results of $\rho_\infty$ correspond to simulations performed in cellular automata with $L = 1,000$.

In general, results are quite similar to those obtained with square lattices, but in the one-dimensional case it is possible to explain, in a simple way, the invasion process and the periodic oscillation that occurs in the steady state, as well as to keep track of the system history.
The patterns present between the cooperative/defective clusters are consequence of the local interaction of the players.
Therefore, observing the time evolution, we can understand the effects of local interactions.
Thus, in the following, we will qualitatively describe the way by the interactions generate the patterns.

In the first time step, $t = 0$, the states of the players are set.
In $t = 1$, the proportion of cooperators decreases, because defectors have higher payoffs, caused by exploiting cooperators.
Exploited players (cooperators) replicate the winners' behavior, in this case defectors.
Thus, these exploited cooperators become defectors, consequently the proportion of defectors increases.
These defectors form defective clusters, creating a border between cooperative and defective clusters.
Notice that these defective clusters are very inconvenient for their members, because the payoff of each player can be lower than the payoff obtained by the players in cooperative clusters due the relations $P_{i}^{(c_i)}(0) > P_{i}^{(c_i)}(1)$ and $P_{i}^{(c)}(\theta) \geq P_{i}^{(c - 1)}(\theta)$, due Eq. (\ref{eq_totalpayoffviz}).
This situation is known as The Tragedy of the Commons.\cite{hardin1968}
If the defectors on the cooperative/defective border have a higher (lower) payoff than the cooperators, then cooperators (defectors) will switch their states to defectors (cooperators) and the defective cluster grows (diminishes), what in evolutionary terms means that defectors (cooperators) are being replicated.
This is the way that invasions of clusters takes place.
Due the connectivity of the players on the border of cooperative/defective clusters the activity of automata occurs on the borders.

\begin{figure}[tb]
\centerline{\psfig{file=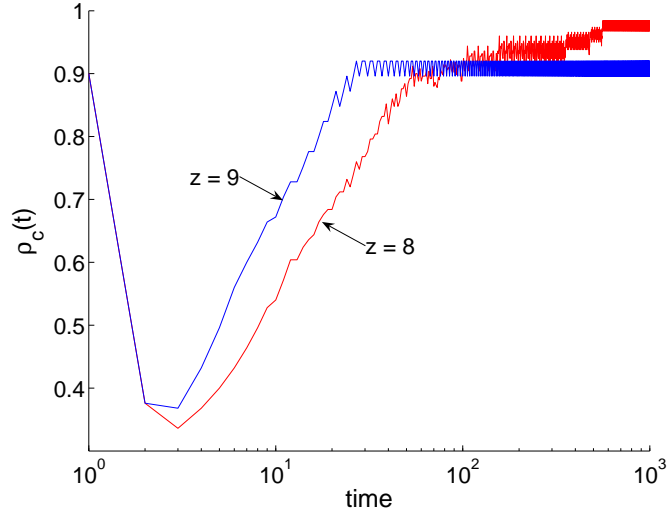,width=0.70\linewidth}}
\vspace*{8pt}
\caption{
Time evolution of the proportion of cooperators in a one-dimensional cellular automata for $z$ symmetric. $\rho_0 = 0.9$, $T = 1.60$, $z = 9$ (with self-interaction) and $z = 8$ (without self-interaction).
\label{fig_plots_time}}
\end{figure}

In Figure \ref{fig_plots_time} the time evolution of the proportion of cooperators, $\rho_c(t)$, is plotted for $z = 9$, with self-interaction, and for $z = 8$, without self-interaction.
One sees that in the initial time steps, $\rho_c(t)$ decays abruptly, increasing quickly in the next steps and finally oscillates around a fixed mean equilibrium value.

In the literature\cite{nowak1992} about PD this behavior is similar, but when $\rho_c(t)$ increases there is an overshoot and then it decreasesaa to the fixed mean equilibrium value.
This indicates that the observed overshoots in the previous systems are due to strong fluctuations of the cooperative/defective clusters that are more susceptible to occur in higher dimensional systems.
In our case, the system converges faster to the asymptotic regime than in square lattice PD for instance.

\begin{figure}[tb]
\begin{center}
\centerline{\psfig{file=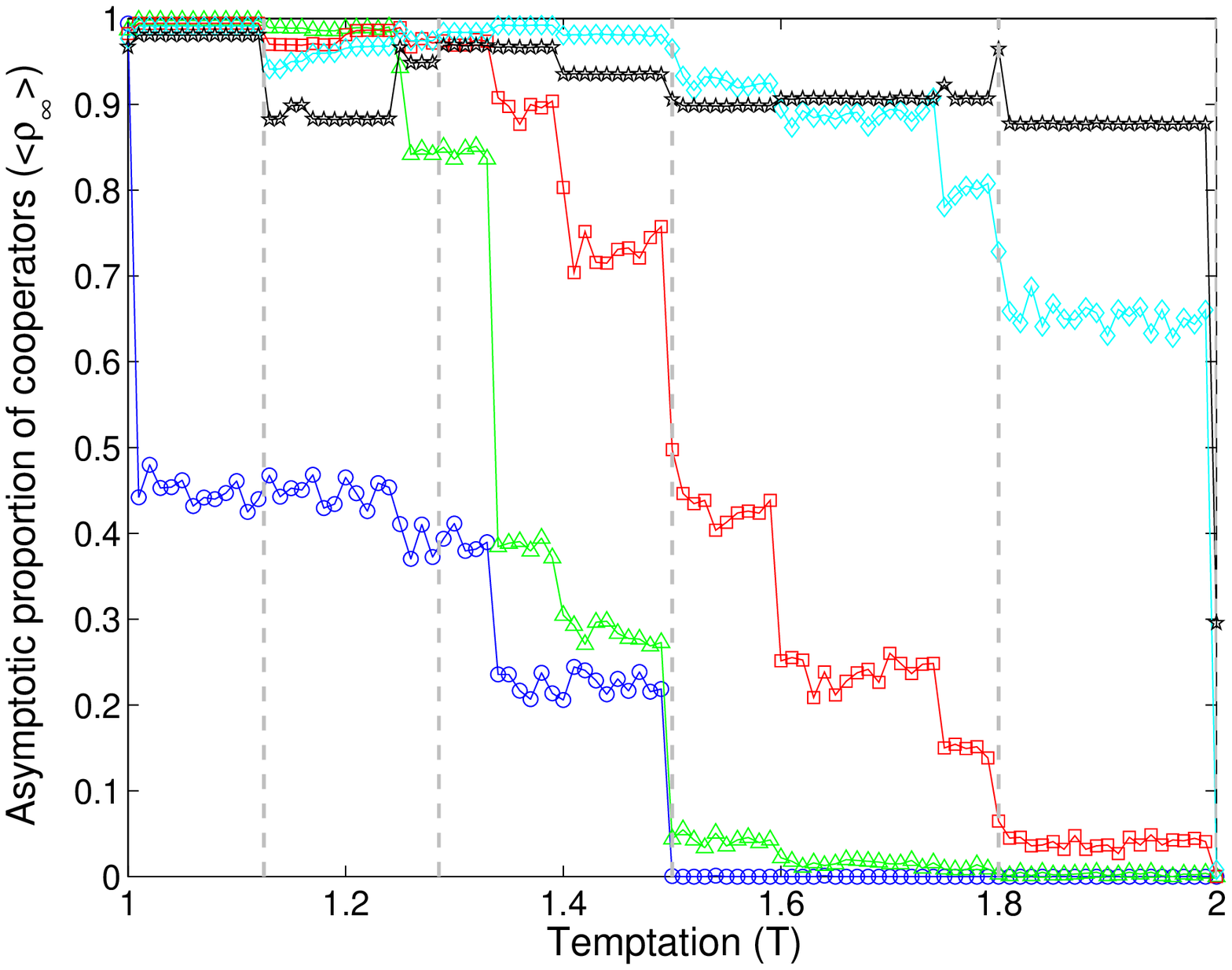,width=.7\linewidth}}
(a)

\centerline{\psfig{file=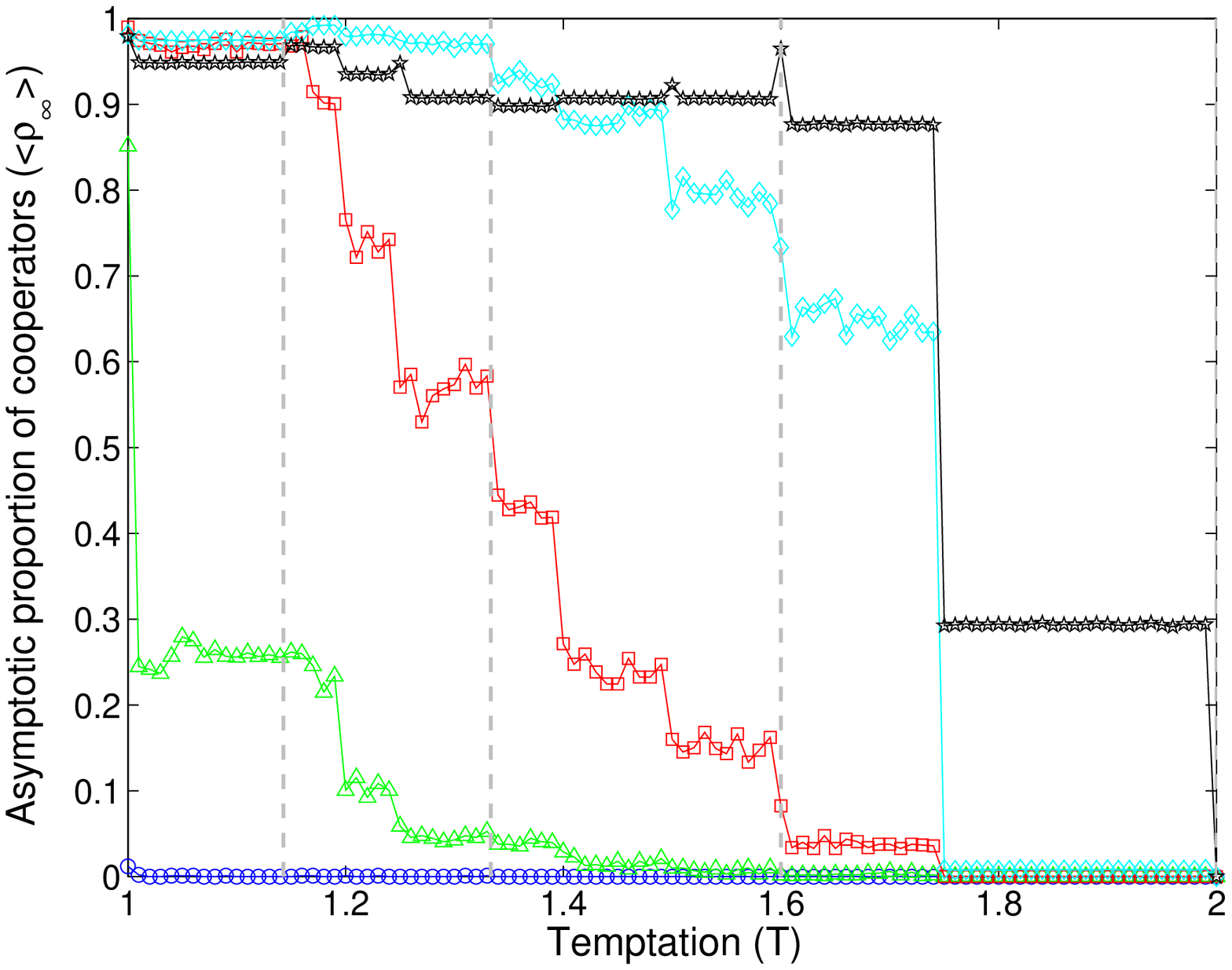,width=.7\linewidth}}
(b)
\end{center}

\vspace*{8pt}
\caption{
Asymptotic proportion of cooperators ($\rho_\infty$) as a function of temptation ($T$) for $z$ symmetric.
Vertical dashed lines denote transition values.
(a) System with $z = 9$ (with self-interaction).
(b) System with $z = 8$ (without self-interaction).
Stars: $\rho_0 = 0.9$; Diamonds: $\rho_0 = 0.7$; Squares: $\rho_0 = 0.5$; Triangles: $\rho_0 = 0.3$; Circle: $\rho_0 = 0.1$.
\label{fig_plots_T_z}}
\end{figure}

In Figures
\ref{fig_plots_T_z}a and
\ref{fig_plots_T_z}b, one checks the relation given by Eq. (\ref{eq_Tc}).
For instance, in Figure
\ref{fig_plots_T_z}a ($z = 9$, $n = 0$ and $m = 4$) there is self-interaction.
The transitions in $\rho_\infty$ can be seen when the parameter $T$ passes through the thresholds of a critical temptation, $T_c$.
For $z = 9$, the coexistence region of cooperation/defection is in the range $9/5 < T < 2$, and the defection region starts at $T = 2$.
Otherwise, if the players do not self-interact, see Figure
\ref{fig_plots_T_z}b ($z = 8$, $n = 1$ and $m = 3$), the region of coexistence of cooperation/defection is in the range $8/5 < T < 5/3$, i.e., close to the middle of the conflict region $1 < T < 2$.
The inclusion of self-interaction implies in higher $\rho_\infty$ values, what means that cooperation prevails when self-interaction is included.
Neglecting self-interaction, $\rho_\infty$ reaches lower values, but steady state and transient regimes are not drastically modified, what validates the results of Ref. \refcite{ricardo2006}.

\begin{figure}[tb]
\centerline{\psfig{file=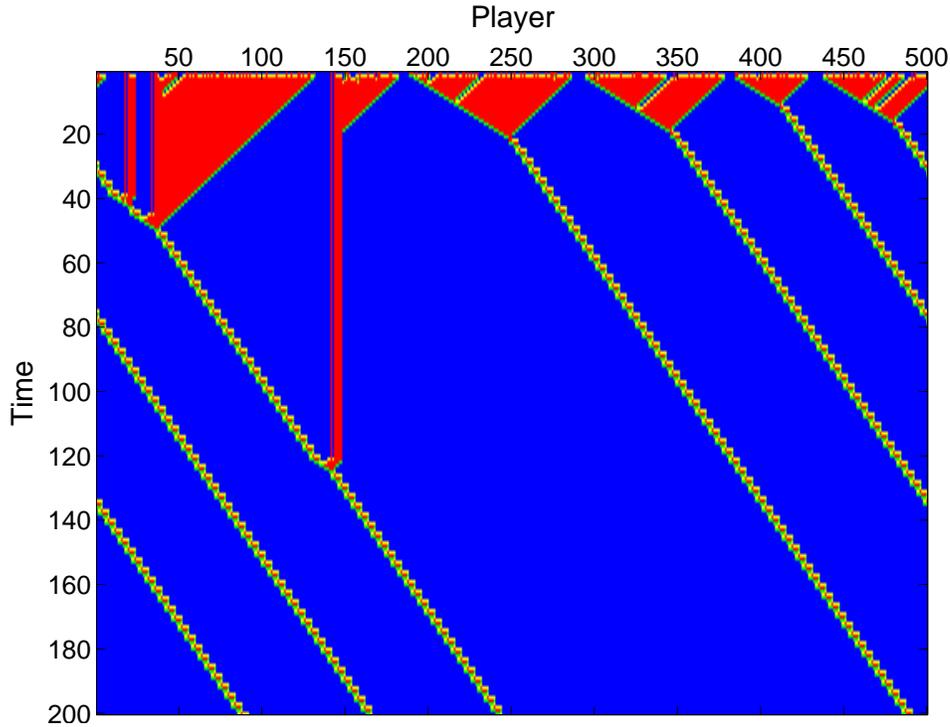,width=1.00\linewidth}}
\vspace*{8pt}
\caption{
Time evolution of the PD in a one-dimensional cellular automata for $z$ asymmetric.
Time evolves from top to bottom.
Each vertical line is the evolution of the states of the $i$ automata.
Blue (dark gray): cooperators; Red (medium gray): defectors; Green (light gray): currently cooperators that were defectors in the previous time step; Yellow (lighter gray): currently defectors that were cooperators in the previous time step.
The parameters in this simulation are $L = 500$, $time = 500$, $T = 1.40$, $\rho_0 = 0.7$, and $z = 7$ (with self-interaction).
\label{fig_automata_01_triangles}}
\end{figure}

\begin{figure}[tb]
\begin{center}
\centerline{\psfig{file=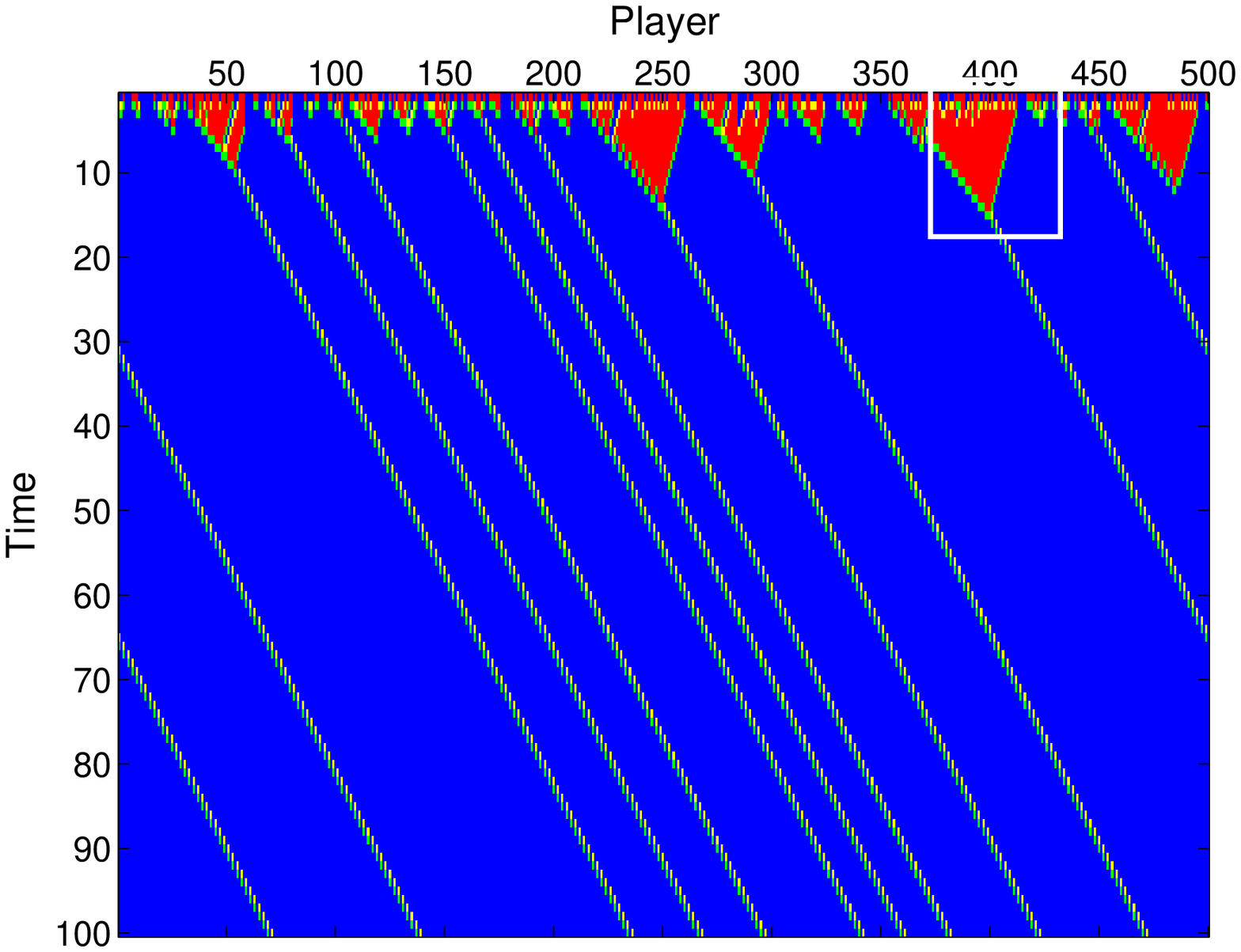,width=0.7\linewidth}}
(a)

\centerline{\psfig{file=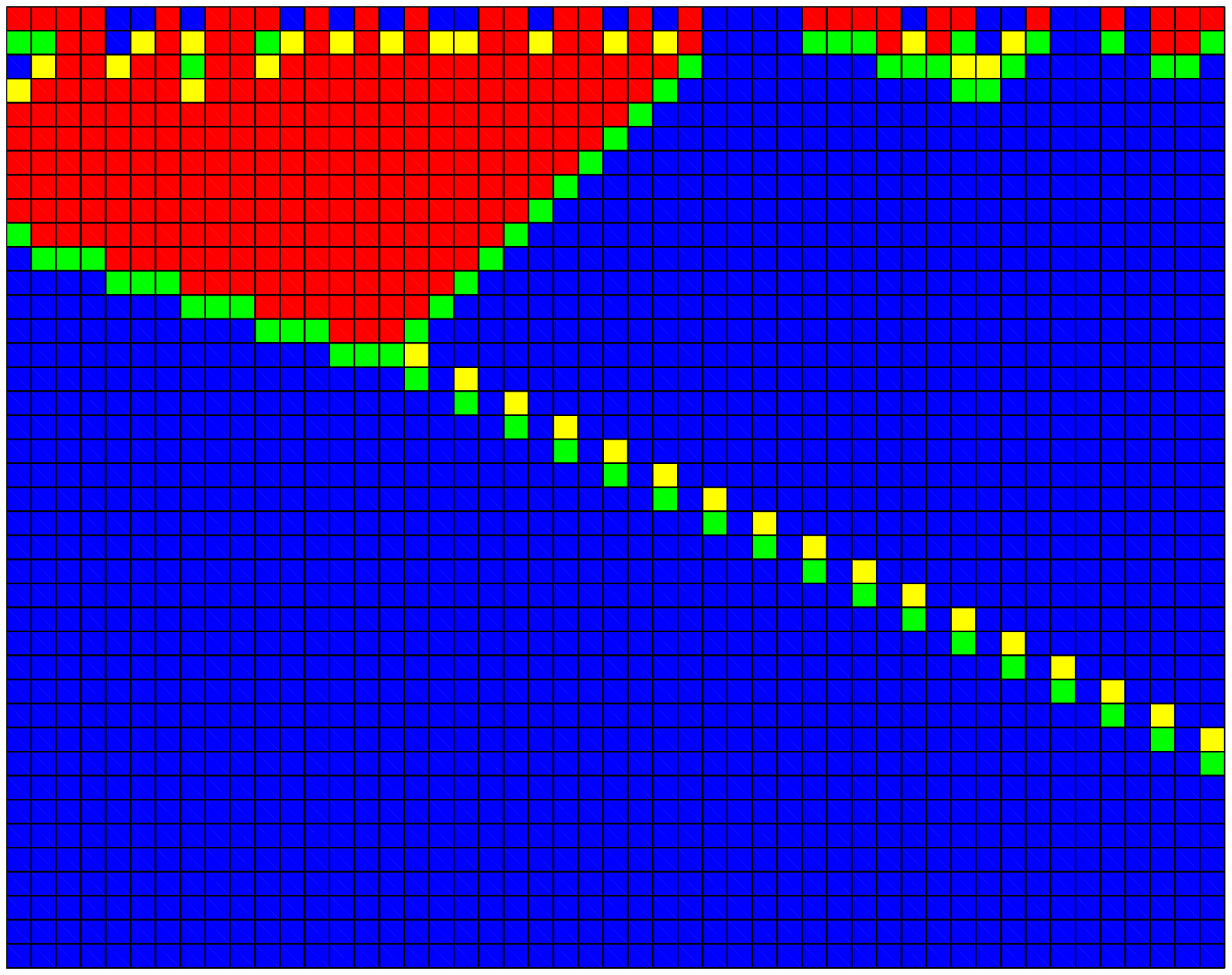,width=0.65\linewidth}}
(b)
\end{center}

\vspace*{8pt}
\caption{
(a) Time evolution of the PD in a one-dimensional cellular automata for $z$ asymmetric.
Time evolves from top to bottom.
Each vertical line is the evolution of the states of the $i$ automata.
Blue (dark gray): cooperators; Red (medium gray): defectors; Green (light gray): currently cooperators that were defectors in the previous time step; Yellow (lighter gray): currently defectors that were cooperators in the previous time step.
(b) Magnification of the regions marked above shows a glider that travels in jumps.
The parameters in this simulation are $L = 500$, $time = 500$, $T = 1.20$, $\rho_0 = 0.5$, and $z = 5$ (with self-interaction).
\label{fig_automata_02_glider_jump}}
\end{figure}

\begin{figure}[tb]
\begin{center}
\centerline{\psfig{file=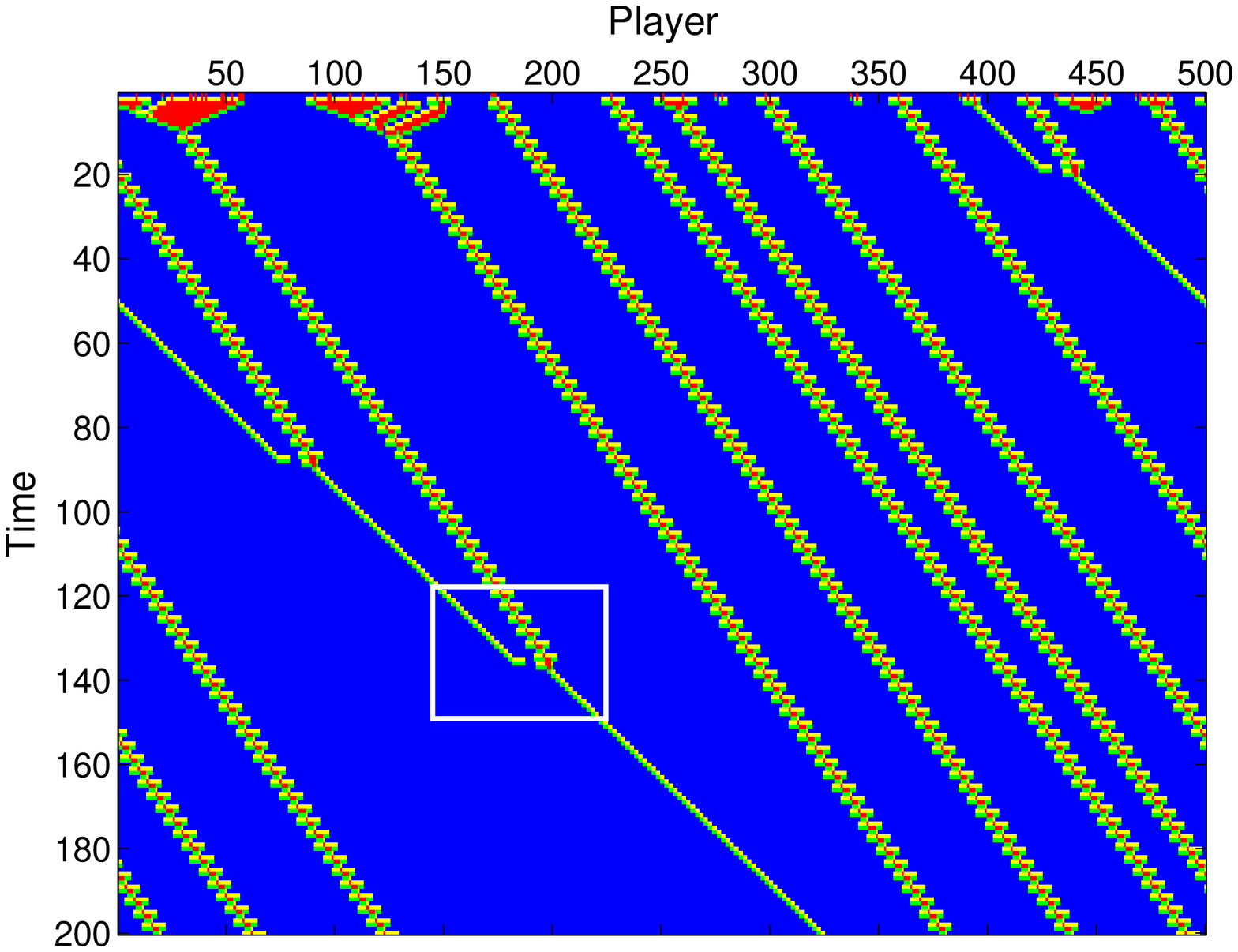,width=0.7\linewidth}}

(a)
\centerline{\psfig{file=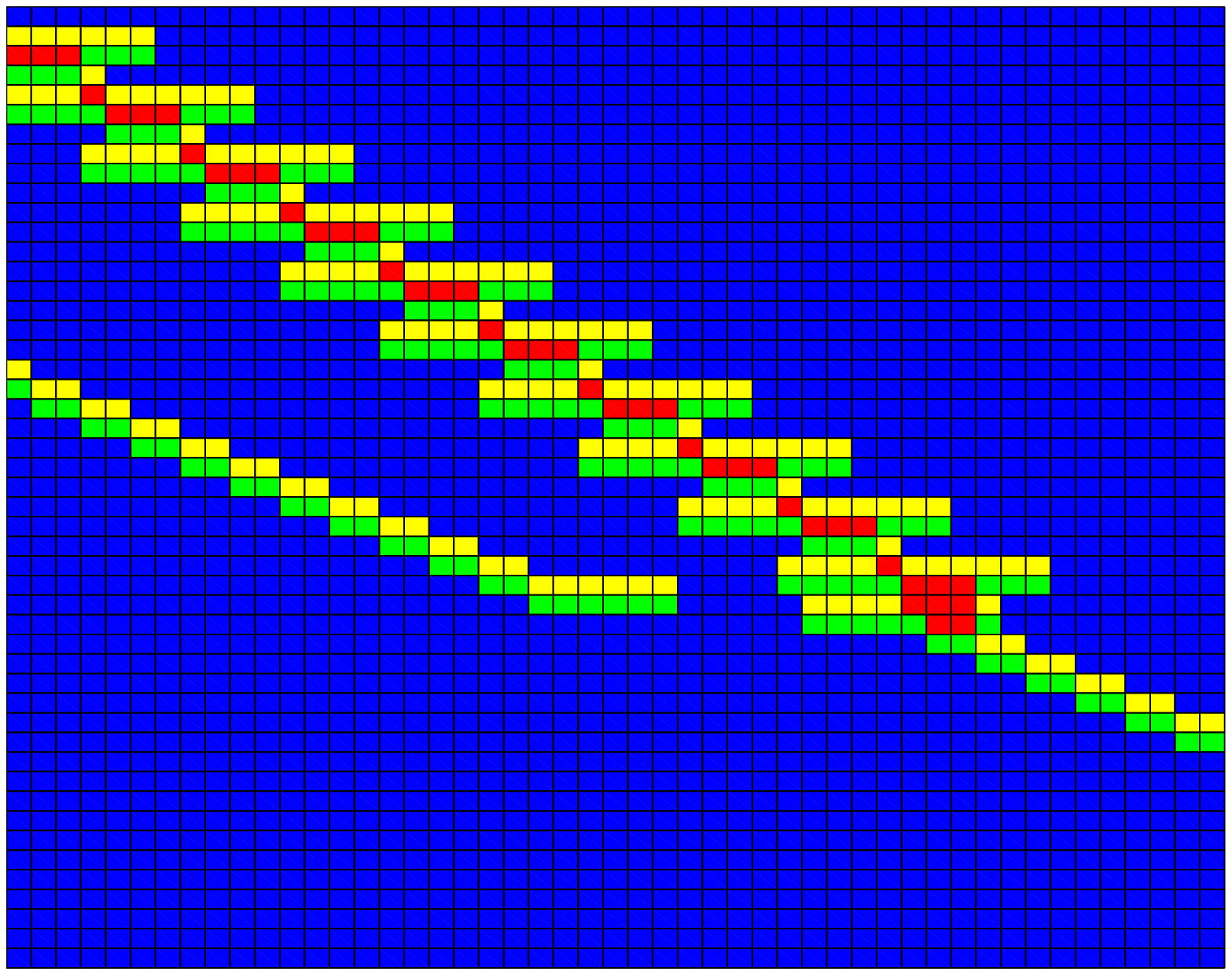,width=0.7\linewidth}}
(b)
\end{center}

\vspace*{8pt}
\caption{
(a) Time evolution of the PD in a one-dimensional cellular automata for $z$ asymmetric.
Time evolves from top to bottom.
Each vertical line is the evolution of the states of the $i$ automata.
Blue (dark gray): cooperators; Red (medium gray): defectors; Green (light gray): currently cooperators that were defectors in the previous time step; Yellow (lighter gray): currently defectors that were cooperators in the previous time step.
(b) Magnification of the region marked above shows the long range interaction of the glider.
The parameters in this simulation are $L = 500$, $time = 500$, $T = 1.20$, $\rho_0 = 0.9$, and $z = 11$ (with self-interaction).
\label{fig_automata_03_interdist_gl-gl}}
\end{figure}

\begin{figure}[tb]
\begin{center}
\centerline{\psfig{file=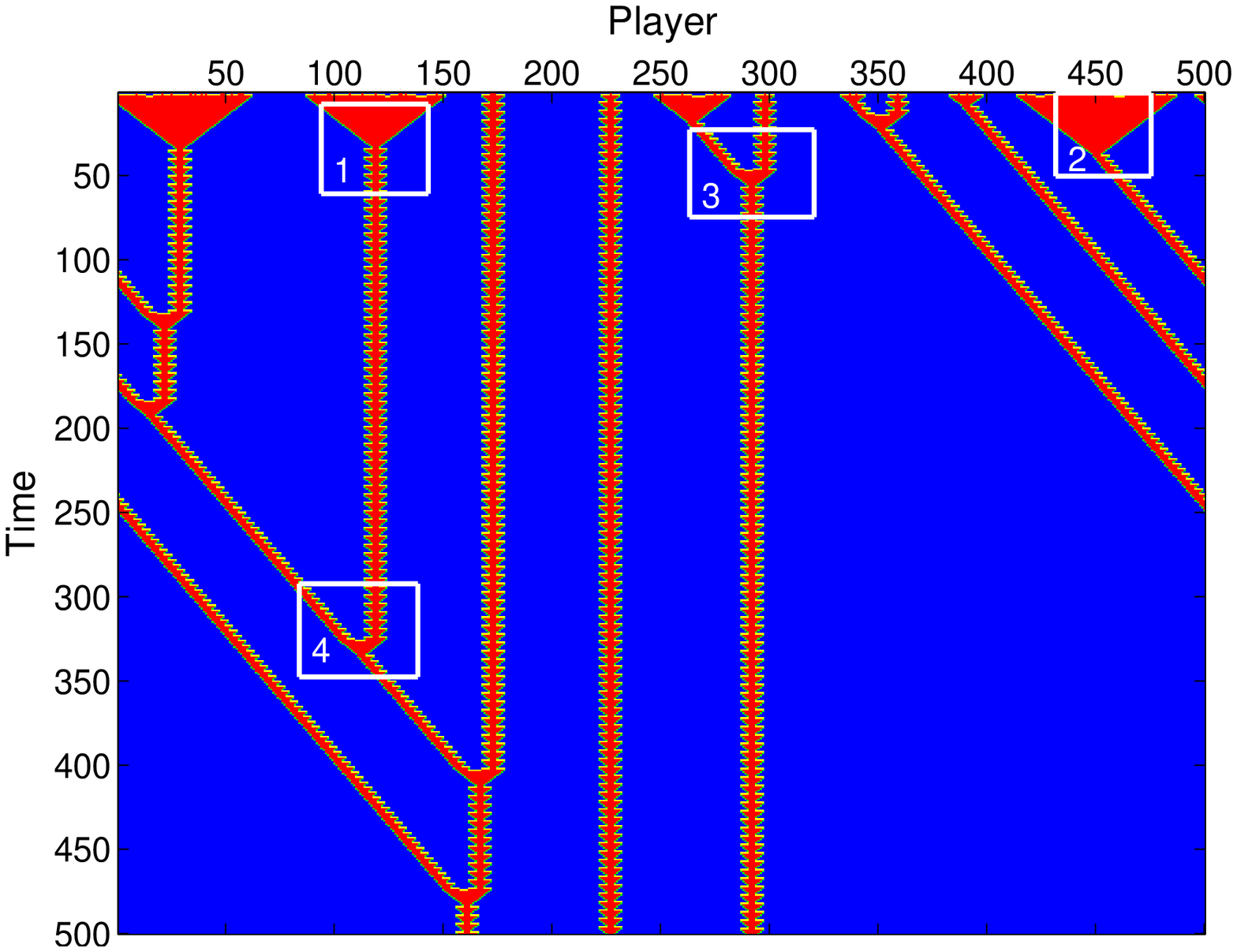,width=0.7\linewidth}}
(a)

\centerline{\psfig{file=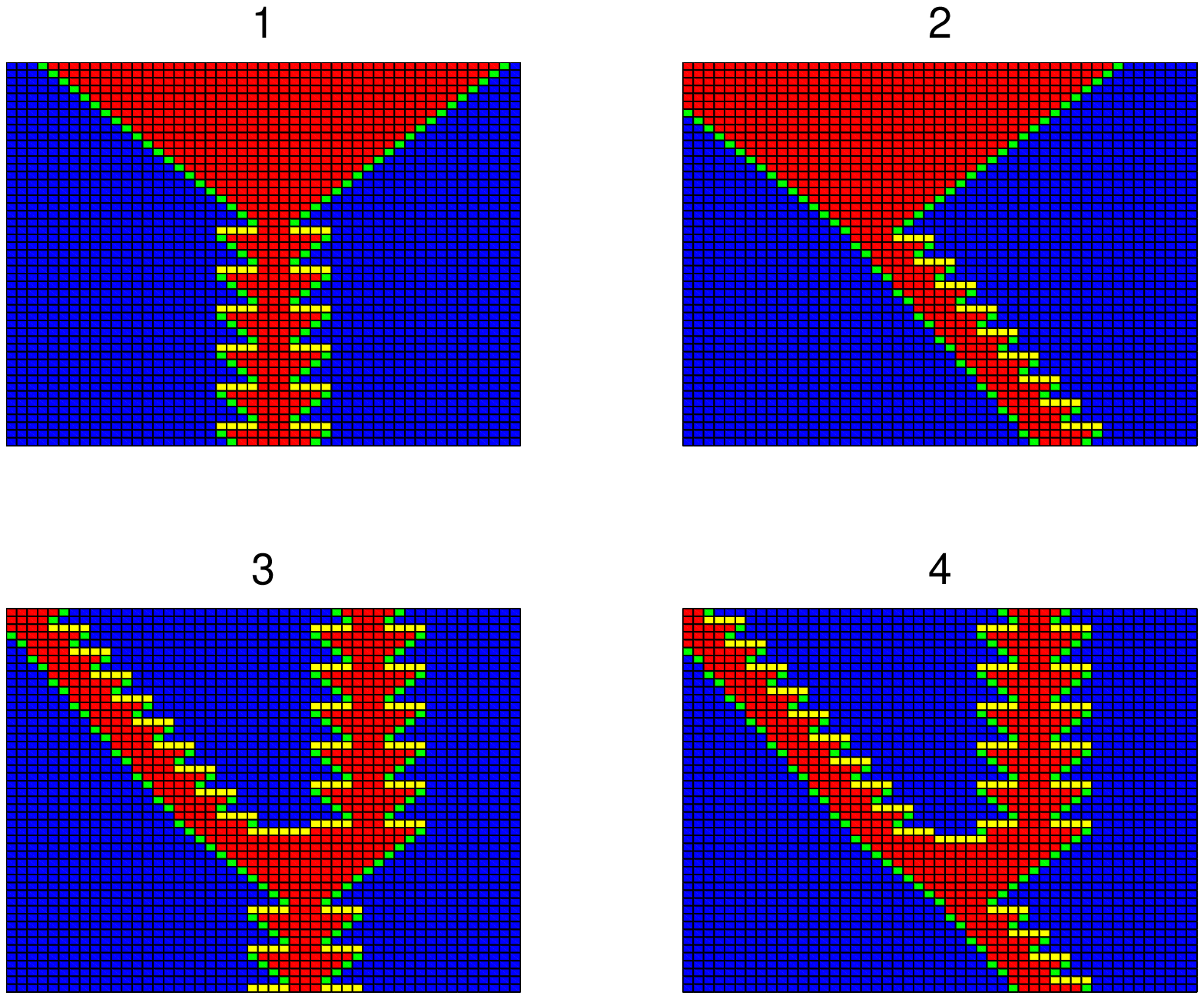,width=0.7\linewidth}}
(b)
\end{center}

\vspace*{8pt}
\caption{
(a) Time evolution of the PD in a one-dimensional cellular automata for $z$ symmetric.
Time evolves from top to bottom.
Each vertical line is the evolution of the states of the $i$ automata.
Blue (dark gray): cooperators; Red (medium gray): defectors; Green (light gray): currently cooperators that were defectors in the previous time step; Yellow (lighter gray): currently defectors that were cooperators in the previous time step.
(b) Magnification of the regions marked above.
(1) finger originated from the end of triangle;
(2) glider originated from the end of triangle;
(3) collision: finger absorbs the glider;
(4) collision: glider absorbs the finger.
The parameters in this simulation are $L = 500$, $time = 500$, $T = 1.60$, $\rho_0 = 0.9$, and $z = 8$ (without self-interaction).
\label{fig_automata_04_collisions}}
\end{figure}

\begin{figure}[tb]
\begin{center}
\centerline{\psfig{file=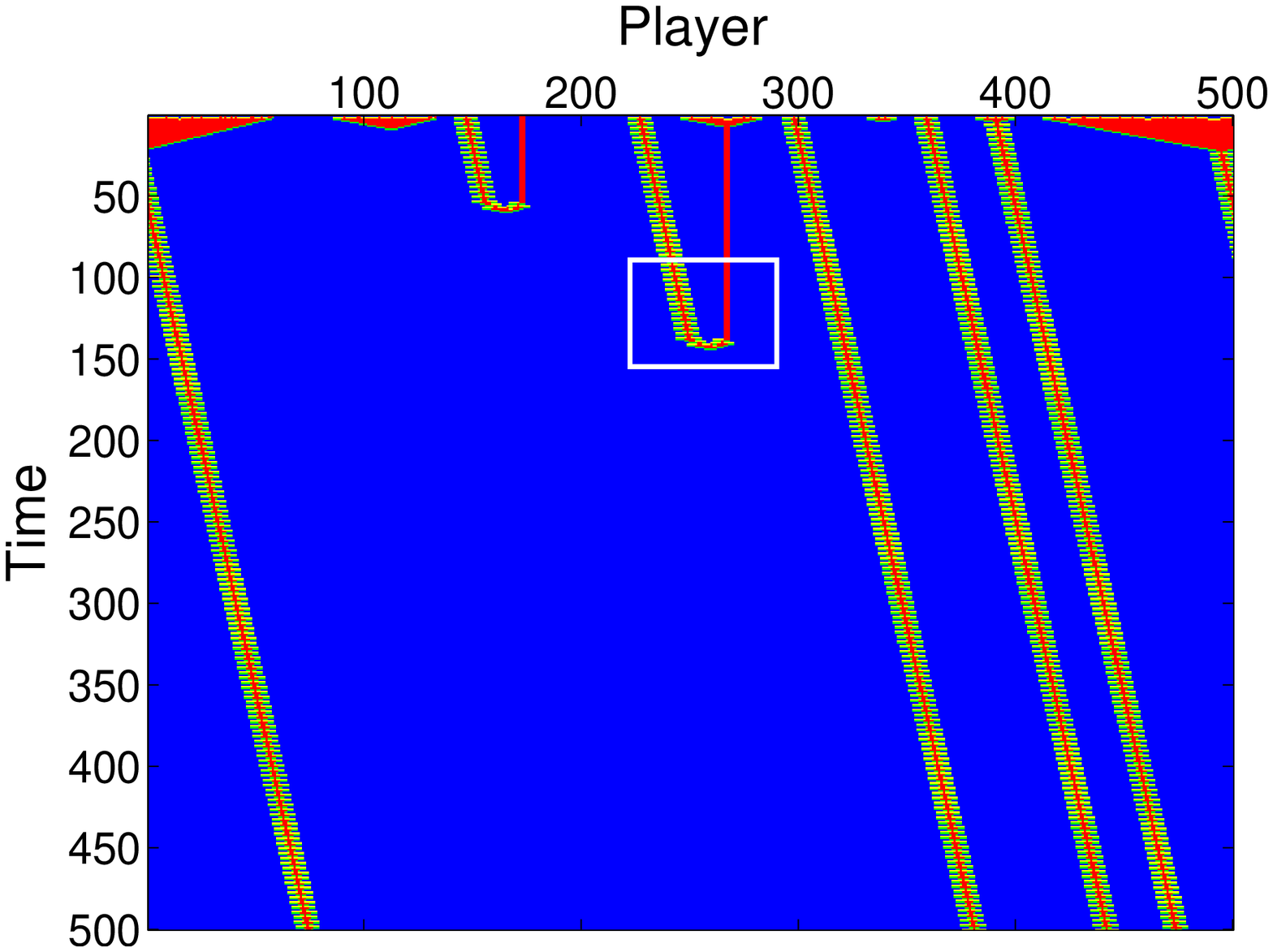,width=0.7\linewidth}}
(a)

\centerline{\psfig{file=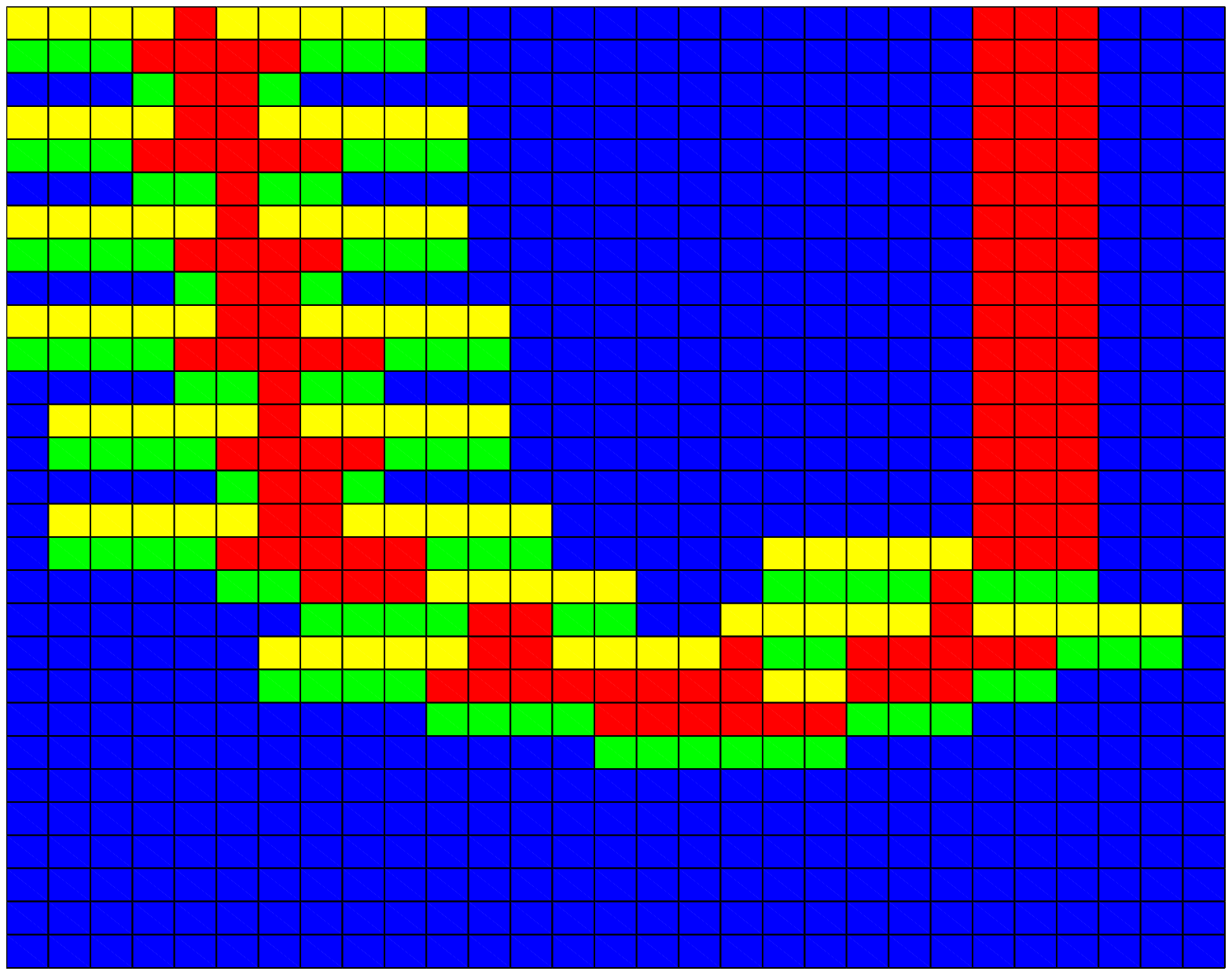,width=0.7\linewidth}}
(b)
\end{center}

\vspace*{8pt}
\caption{
(a) Time evolution of the PD in a one-dimensional cellular automata for $z$ symmetric.
Time evolves from top to bottom.
Each vertical line is the evolution of the states of the $i$ automata.
Blue (dark gray): cooperators; Red (medium gray): defectors; Green (light gray): currently cooperators that were defectors in the previous time step; Yellow (lighter gray): currently defectors that were cooperators in the previous time step.
(b) Magnification of the region marked above shows the annihilation of two colliding gliders.
The parameters in this simulation are $L = 500$, $time = 500$, $T = 1.20$, $\rho_0 = 0.9$, and $z = 10$ (with self-interaction).
\label{fig_automata_05_coll_ext}}
\end{figure}

In the initial steps after the formation of cooperative/defective clusters, they form patterns like triangles.
Figures
\ref{fig_automata_01_triangles},
\ref{fig_automata_02_glider_jump}a,
\ref{fig_automata_03_interdist_gl-gl},
\ref{fig_automata_04_collisions}a, and 
\ref{fig_automata_05_coll_ext}a,
show some examples of the dynamics of patterns formation.
The bases of these triangles are the initial cooperative/defective clusters, and the triangles slopes depend on the parameters set in the simulation.
As time evolves, these triangles becomes narrower, and they can eventually be extinguished.
If the triangles do not extinguish, they can generate more persistent structures: the fingers and the gliders.

A finger is a pattern that extends itself across the system as a straight line originated in the end of the triangle.
Its edge may be flat or rough.
Rough edges can be complex, a saw-tooth for example, presenting important features as periodicity and reflection.

In its turn, a glider is a pattern that travels across the system and it can collide with others patterns.
It can also be used to transmit information over long distances.\cite{wolfram1983}
It seems a sloped finger, and was originally described in the Game of Life\cite{gardner1970}, where it represented a spaceship that travels diagonally.

The origin of the finger, glider and also the glider's slope depends on the number of defectors in the end of the triangle being essentially determined by $z$ and $T$.

Some gliders can been observed in Figures
\ref{fig_automata_01_triangles},
\ref{fig_automata_02_glider_jump}a,
\ref{fig_automata_03_interdist_gl-gl},
\ref{fig_automata_04_collisions}a, and 
\ref{fig_automata_05_coll_ext}a,.
Notice that gliders can coexist with triangles as shown in Figure
\ref{fig_automata_01_triangles}.
In this figure one can also see cooperators gliders that travel inside the defectors triangles, but in the mirrored direction of defectors gliders.
Also, observe that a glider can propagate continuously in spatio-temporal evolution or by jumps as shown in Figure \ref{fig_automata_02_glider_jump}b, that is a magnification of the area marked by a square in Figure \ref{fig_automata_02_glider_jump}a.
The propagation in jumps occurs only for asymmetric neighborhood, symmetric neighborhood generates continuous patterns.

There are several types of intersections among these patterns.
For instance, we mention collisions: glider-triangle, glider-finger and glider-glider.
Fingers do not collide with fingers because they evolve parallel each other.
Glider-glider collision is present only with asymmetric neighborhood, for symmetric neighborhood the gliders have the same slope, so they do not collide.
Observe the collisions of:
glider-triangle in Figure
\ref{fig_automata_01_triangles} around player 50, time 50;
glider-finger in Figures
\ref{fig_automata_04_collisions}b-3;
and glider-glider in Figure
\ref{fig_automata_05_coll_ext}b.
When a glider collides with another pattern there are several possible outcomes:
i) it can be absorbed (Figure
\ref{fig_automata_04_collisions}b-3),
ii) it absorbs the other pattern (Figure
\ref{fig_automata_04_collisions}b-4),
iii) it interacts with the pattern yielding a different pattern from the original ones or
iv) they are annihilated (Figure
\ref{fig_automata_05_coll_ext}).
If the glider is absorbed, or it absorbs the other pattern, the emergent pattern can keep its original direction or be drifted, if it is a finger; or it can keep or change its original slope, if it is a glider.

The glider can also interact with other patterns without collision among them, in the case of asymmetric neighborhood.
This is the case of a long range interaction around it.
Figure
\ref{fig_automata_03_interdist_gl-gl}b exemplifies the long range interaction of two gliders.

The defective gliders drift to the right hand side because the player states are updated from the left to the right hand side.
If they were updated in the other sense the drift would be to left hand side.

In brief, three scenarios come out as result of these local interactions.
Firstly, defective clusters grow until the completely extinction of cooperators, reaching a defective phase, $\rho_\infty = 0$.
Secondly, defective clusters grow and dominate the system, but cooperators are not extinguished, resulting in a defective phase as well, $0 < \rho_\infty < 0.5$.
And finally, the defective clusters are invaded by cooperators resulting in a cooperative phase, $\rho_\infty > 0.5$.

Numerical simulations of the PD have shown that in principle, patterns change if the initial configuration of the system is modified.
But this does not necessarily occur when the parameters are modified.
For a given configuration of $z$ and $\rho_0$ there are ranges of $T$ that the patterns do not change.
In these ranges, only the players' payoff are modified, what is expected if we remember the regions between the transitions values, $T_c$.
Moreover, there are also regimes where only one parameter dominates the whole dynamic, like in the case of high values of $z$ ($> 20$).
The oscillating behavior of $\rho_\infty$ can now be easily understood watching the saw-tooth in the patterns emerged.
In previous works, this fluctuation were attributed to the spatial patterns oscillations.
Finally we point out that the patterns obtained in the PD in the one-dimensional cellular automata belong to classes 1, 2 and 4 of the classes proposed in Ref. \refcite{wolfram1983}.

\section{Conclusion} \label{conclusao}
The one-dimensional cellular automata with each cell being a player that plays the Prisoner's Dilemma, with his/her $z$ neighboring cells (including self-interaction for odd values of $z$) are interesting in several aspects.
It allows us to retrieve all the previous results obtained for regular lattices in $d$ dimensions.
Since in regular lattices the process of invasion of a cooperative (defective) cluster by defectors (cooperators) occurs in the corner players (the ones with the least cooperative neighbors), spatial geometry allows large fluctuations leading to bumps (overshoot) in transient times.
In our system, these bumps (overshoots) do not occur.
The invasion process occurs at most in two sites (extremes of the cooperative/defective cluster).
Convergence to stationary values is faster in the one-dimensional automata than in an arbitrary regular lattices.
The other advantage of the proposed one-dimensional automata is that one can keep track of the player history.
Following the players behavior along time, give rises to several interesting patterns, such as glider and fingers.
The thickness and the slope of these structures are essentially given by $z$ and $T$.
Cooperative and defective clusters can now be interpreted as particles that interact one with the other, where they can be extinguished or coexist.
The comprehension of the collision processes furnishes a new approach to characterize the rich phase diagram of Spatial Prisoner's Dilemma.

\section*{Acknowledgments}
M. A. P. would like to thank CAPES for the fellowship.
A. L. E. would like to thank CNPq (06/60333-0) for the fellowship and FAPESP and MCT/CNPq Fundo Setorial de Infra-Estrutura for the financial support.
A. S. M. acknowledges the agencies CNPq (305527/2004-5) and FAPESP (2005/02408-0) for support.

\end{document}